\documentclass[aps,prb,twocolumn,reprint,amsmath,amssymb,superscriptaddress]{revtex4-2}

\usepackage{amsmath}
\usepackage{graphicx}
\usepackage{dcolumn}
\usepackage{bm}
\usepackage{multirow}
\usepackage{subfigure}
\usepackage{booktabs}
\usepackage{hhline}  
\usepackage{float}
\usepackage{braket}
\usepackage[colorlinks,linkcolor=blue,anchorcolor=blue,citecolor=blue,urlcolor=blue]{hyperref}
\usepackage[normalem]{ulem}
\usepackage{hyperref}

\begin{document}


\title{Stochastic path-integral approach for predicting the superconducting temperatures of anharmonic solids}


\author{Haoran Chen}
\affiliation{International Center for Quantum Materials, Peking University, Beijing 100871, China}

\author{Junren Shi}
\email{junrenshi@pku.edu.cn}
\affiliation{International Center for Quantum Materials, Peking University, Beijing 100871, China}
\affiliation{Collaborative Innovation Center of Quantum Matter, Beijing 100871, China}


\date{\today}

\begin{abstract}
We develop a stochastic path-integral approach for predicting the superconducting transition temperatures of anharmonic solids.  By defining generalized Bloch basis, we generalize the formalism of the stochastic path-integral approach, which is originally developed for liquid systems.  We implement the formalism for \textit{ab initio} calculations using the projector augmented-wave method, and apply the implementation to estimate the superconducting transition temperatures of metallic deuterium and hydrogen sulfide.  For metallic deuterium, which is approximately harmonic, our result coincides well with that obtained from the standard approach based on the harmonic approximation and the density functional perturbation theory. For hydrogen sulfide, we find that anharmonicity strongly suppresses the predicted superconducting transition temperature.  Compared to the self-consistent harmonic approximation approach, our approach yields a transition temperature closer to the experimentally observed one. 
\end{abstract}

\maketitle

\section{Introduction}
Atomic vibrations mediate attractive interactions between electrons, and induce superconductivity in solids. For most solids, harmonic description of atomic vibrations is satisfactory~\cite{grimvall_electron-phonon_1981}. 
Based on the description, first-principles approaches like density functional perturbation theory (DFPT) are developed, and achieve great successes in calculating properties related to electron-phonon coupling (EPC) and predicting superconducting transition temperatures ($T_c$)~\cite{giustino_electron-phonon_2017,baroni_phonons_2001}.
For systems containing light atoms like hydrogen, however, vibrating amplitudes can be so large that the harmonic description breaks down.
Hydrides, many of which are high-$T_{c}$ conventional superconductors, are such examples~\cite{stritzker_superconductivity_1972,schirber_concentration_1974,drozdov_conventional_2015,drozdov_superconductivity_2019}. In these compounds, hydrogen atoms tend to have large vibrating amplitudes, introducing strong anharmonicity. Moreover, driven by quantum fluctuations, they could tunnel across potential barriers, resulting in the super-ionic phase in some cases~\cite{wang_quantum_2021,liu_dynamics_2018}. These effects can lead to corrections to phonon spectra, EPC, and the stabilization of crystal structures~\cite{errea_quantum_2016,errea_quantum_2020}. It is shown that the corrections could either enhance or suppress predicted $T_{c}$~\cite{errea_first-principles_2013,errea_quantum_2016,errea_quantum_2020,liu_dynamics_2018}. The conventional harmonic approaches are thus unreliable in predicting $T_{c}$ for hydrides. A systematic approach fully taking account of the effects of anharmonicity and quantum fluctuations is therefore needed.

A well-established non-perturbative approach for dealing with these effects is the stochastic self-consistent harmonic approximation (SSCHA)~\cite{werthamer_self-consistent_1970,souvatzis_entropy_2008,errea_anharmonic_2014}, which employs a trial harmonic potential to model an anharmonic system. With the harmonic potential, one can determine an effective dynamical matrix, and thus the spectrum and eigen-modes of phonons.  EPC matrix elements are then calculated in the same way as in the conventional harmonic approach using the effective phonons~\cite{errea_first-principles_2013}. While the approach yields reasonable results for many systems, it is nevertheless based on an uncontrolled approximation.  For systems in which ions have large vibrating amplitudes or even move around freely,  the validity of the underlying assumption that the system can be modeled by an effective harmonic one is questionable.

Recently, Liu \textit{et al.} proposed a stochastic path-integral approach (SPIA) for predicting $T_c$ of metallic liquids~\cite{liu_superconducting_2020}. The approach is based on the \textit{ab initio} path integral molecular dynamics (PIMD) technique, and directly determines the effective attractive interaction between electrons by analyzing the fluctuations of electron-ion-scattering matrices. The approach is based on a set of rigorous relations without resorting to defining an effective harmonic system.  More importantly, the approach makes no assumptions on the nature of ion motion, and therefore can be applied to general systems including anharmonic solids.  More recently, Chen \textit{et al.} develop an implementation of the approach based on the density functional theory (DFT) and the projector augmented-wave (PAW) method~\cite{chen_first-principles_2021}.
While the implementation is only applicable for liquids, it solves a key issue of applying the SPIA approach for more general systems, i.~e., determining electron-ion scattering amplitudes accurately and efficiently.

In this paper, we extend the approach for anharmonic solids. In a solid, the continuous translational invariance breaks down. 
Electrons experience a periodical crystal potential which gives rise to Bloch states, between which Cooper pairing occurs. As a result, physical quantities like effective interactions need to be expanded in the Bloch states instead of plane waves as in liquids.
To this end, we define generalized Bloch states for an anharmonic solid by introducing an effective Hamiltonian, and generalize the formalism of SPIA accordingly. We implement the formalism by using the PAW method. A number of issues associated with the implementation, such as transformations of spherical waves at different centers and the over-sampling in a time-domain, are also solved.  As tests, we apply the implementation to metallic deuterium, which is approximately harmonic, and hydrogen sulfide H$_3$S, which is strongly anharmonic. For the former, our results coincide well with those obtained from the conventional approach based on the harmonic approximation and DFPT. For the latter, we find that anharmonic effects strongly suppress predicted $T_c$. Compared with the SSCHA approach, our approach yields $T_{c}$ closer to the experimentally observed one.

The remainder of the paper is organized as follows.  In Sec.~\ref{SPIA}, we develop the formalism of the SPIA for solids.  We then discuss our \textit{ab initio} implementation of the formalism using the PAW method in Sec.~\ref{G_PAW}.  In Sec.~\ref{Hydrogen} and~\ref{H3S} we apply the implementation to metallic deuterium D and hydrogen sulfide H$_3$S, and compare results with previous calculations and experiments. Finally, we summarize our results in Sec.~\ref{Summary}.  Miscellaneous details of the theory and tests are presented in Appendices.

\section{Stochastic path-integral approach for solids}\label{SPIA}
The SPIA is a non-perturbative approach for predicting $T_c$ of EPC-induced superconductors~\cite{liu_superconducting_2020}. The central quantity to be determined is an effective interaction between electrons induced by fluctuating ionic fields. In the approach, the fluctuation is sampled by employing the \textit{ab initio} PIMD technique~\cite{chandler_exploiting_1981,marx_ab_1996}, which provides a set of ion configurations $\bm{R}(\tau)\equiv\{\bm{R}_a(\tau)\}$, where $\bm{R}_a(\tau)$ denotes the position of ion $a$ at an imaginary time $\tau$.  One can obtain the effective interaction by determining $T$ matrices of electron-ion scatterings and analyzing their fluctuation.

To determine the $T$ matrix for a given ion configuration $\bm{R}(\tau)$, we first determine the  Green's function $\hat{\mathcal{G}}[\bm{R}(\tau)]$ of an electron subjected to an ionic field with respect to the ion configuration~\cite{liu_superconducting_2020,chen_first-principles_2021,zhang_nonperturbative_2022}. Its average $\hat{\bar{\mathcal{G}}}=\langle\hat{\mathcal{G}}[\bm{R}(\tau)]\rangle_C$ over the sampled ion configurations gives the physical Green's function of an electron in the system.  The scattering $T$ matrix can then be determined by applying the identity
\begin{eqnarray}\label{T_v2}
	\hat{\mathcal{T}}[\bm{R}(\tau)]
	=\hbar\hat{\bar{\mathcal{G}}}^{-1}
	\left(\hat{\mathcal{G}}[\bm{R}(\tau)]-\hat{\bar{\mathcal{G}}}\right)
	\hat{\bar{\mathcal{G}}}^{-1}.
\end{eqnarray}

The average Green's function $\hat{\bar{\mathcal{G}}}$ describes how electrons propagate.  It is natural to define a set of generalized Bloch modes which can propagate in the system without being scattered (but may be damped).  It means that the set of the generalized Bloch modes should approximately diagonalize $\hat{\bar{\mathcal{G}}}$.  In ordinary solids, Bloch states are determined by assuming that all ions are fixed at their equilibrium positions. However, for anharmonic solids, the vibrating amplitudes of ions can be large, and the effective crystal potential experienced by electrons may deviate significantly from the one generated by fixed ions. 
In this case, we can define an effective Hamiltonian
\begin{eqnarray}\label{Heff}
	\hat{H}_{\mathrm{eff}}=-
	\frac{\hbar}{2}\left[\hat{\bar{\mathcal{G}}}^{-1}+\mathrm{h.c.}\right].
\end{eqnarray}
The Hamiltonian reduces to the ordinary Bloch Hamiltonian for harmonic solids.  For anharmonic or superionic solids, $\hat{H}_{\mathrm{eff}}$ provides a generalized Hamiltonian from which one can determine generalized Bloch states. Using them as basis, $\hat{\bar{\mathcal{G}}}$ is diagonal approximately:
\begin{eqnarray}
	\bar{\mathcal{G}}_{11'}\approx\bar{\mathcal{G}}_1\delta_{11'},
\end{eqnarray}
where $1\equiv(n\bm{k},\omega_j)$ denotes the index of the generalized Bloch states, including a quasi-wave-vector $\bm{k}$, a band index $n$ and the Fermion Matsubara frequency $\omega_j$~\cite{mahan_many-particle_2000}, and $1'$ denotes another set of these parameters. We note that the effective Hamiltonian defined in Eq.~(\ref{Heff}) and the Bloch states actually depend on the Matsubara frequency. Fortunately, the dependence is usually weak in real systems.  We can set it to a large frequency so that the quasi-static approximation can be applied for determining the Green's functions~\cite{liu_superconducting_2020,chen_first-principles_2021,zhang_nonperturbative_2022}.
On the other hand, the anti-Hermitian part of $\hat{\bar{\mathcal{G}}}^{-1}$ indicates the presence of damping for propagating electrons. It is also small in real materials, and can be ignored.

With the generalized Bloch states as basis, the effective interaction $\hat{W}$ can be determined by solving the Bethe-Salpeter equation
\begin{eqnarray}\label{BSE_v2}
	W_{11'}=\Gamma_{11'} + \frac{1}{\hbar^2\beta}\sum_2 W_{12}|\bar{\mathcal{G}}_2|^2\Gamma_{21'},
\end{eqnarray}
where $\Gamma_{11'}$ is the fluctuation of the scattering $T$ matrices:
\begin{eqnarray}\label{Gamma_v2}
	\Gamma_{11'}
	=-\beta\left\langle|\mathcal{T}_{11'}[\bm{R}(\tau)]|^2\right\rangle_C.
\end{eqnarray}
The derivation of the equation is identical to that presented in Ref.~\onlinecite{liu_superconducting_2020} with the state indices $1,1'$ being interpreted as indices for the generalized Bloch states instead of plane waves.

The attractive interaction induces the Cooper instability of electrons. In conventional superconductors, we assume that an electron of state $(n,\bm{k},\uparrow)$ is paired with another electron of the time-reversal state $(n,-\bm{k},\downarrow)$, where the arrows denote the directions of electron spins.  Following the same analysis as in Sec.~II~B~2 of Ref.~\onlinecite{liu_superconducting_2020}, we can obtain the linearized Eliashberg equation for solids:
\begin{multline}\label{Eliashberg_v2}
	\rho\Delta_{n\bm{k}j}=
	\sum_{n'\bm{k'},j'}\bigg[
	-\frac{\hbar\beta}{\pi}\left|\tilde{\omega}_{n\bm{k}}(j)\right|\delta_{n\bm{k}j,n'\bm{k'}j'}\\
	-W_{n\bm{k},n'\bm{k'}}(j-j')\delta(\varepsilon_{n'\bm{k'}}-\varepsilon_F)
	\bigg]\Delta_{n'\bm{k'}j'},
\end{multline}
with $\tilde{\omega}_{n\bm{k}}(j)=\omega_j-\mathrm{Im}\bar{\Sigma}(n\bm{k},i\omega_j)/\hbar$, $ j\in Z$, and $\mathrm{Im}\bar{\Sigma}$ is the imaginary part of the self energy induced by EPC and can be determined by applying the generalized optical theorem~\cite{liu_superconducting_2020}; $\varepsilon_{n\bm{k}}$ is the eigen-energy obtained from $\hat{H}_{\mathrm{eff}}$, and $\varepsilon_F$ is the Fermi energy. Equation~(\ref{Eliashberg_v2}) is an eigen-equation with $\rho$ being its eigenvalue, and can be solved in the subspace of states $(n\bm{k})$ restricted to the Fermi surface. The emergence of a positive eigenvalue indicates the instability towards forming Cooper pairs, and therefore a superconducting state~\cite{liu_superconducting_2020,chen_first-principles_2021,allen_transition_1975}.  We note that $\Delta_{n\bm{k}j}$, which is proportional to the pairing amplitude of Cooper pairs, depends on the band index $n$ and quasi-wave-vector $\bm{k}$, as well as the Matsubara frequency index $j$. This is different from liquid systems, for which $\Delta$ only depends on $j$, and not on $n$ and $\bm{k}$ because a liquid system has no Bloch bands and is isotropic.  It complicates the solution of the Eliashberg equation~(\ref{Eliashberg_v2}).
 
It is usually satisfactory to apply the isotropic approximation when solving Eq.~(\ref{Eliashberg_v2}).  This is to assume that $\Delta_{n\bm{k}j}$ is independent of $n$ and $\bm{k}$, and depends only on $j$.  Under the approximation, one replaces the effective interaction $\hat{W}$, which is anisotropic in general, with an isotropic one. We define a set of EPC interaction parameters as the Fermi-surface-average of $W$~\cite{allen_transition_1975}:
\begin{multline}\label{lambda_avg}
	\lambda(j-j')=-\frac{1}{N(\varepsilon_F)}\sum_{n\bm{k},n'\bm{k'}}
	W_{n\bm{k},n'\bm{k'}}(j-j')\\
	\times\delta(\varepsilon_{n\bm{k}}-\varepsilon_F)\delta(\varepsilon_{n'\bm{k'}}-\varepsilon_F),
\end{multline}
where $N(\varepsilon_F)$ is the electron density of states (DOS) at the Fermi level.  To include the effect of the Coulomb interaction between electrons, we introduce the Morel-Anderson pseudopotential $\mu^{*}$~\cite{morel_calculation_1962}.  Combining all, we obtain the usual linearized Eliashberg equation for isotropic systems~\cite{liu_superconducting_2020,allen_transition_1975}, with interaction parameters defined by Eq.~(\ref{lambda_avg}). This will be the equation for determining $T_c$ in this work.

There may be cases for which the isotropic approximation is not satisfactory.  It is well known that MgB$_2$ is a multi-gap superconductor, the pairing amplitude of which depends on the band index $n$~\cite{liu_beyond_2001,golubov_specific_2002}. EPC superconductors with paring amplitudes strongly depending on $\bm{k}$ are also predicted in theory~\cite{abrikosov_nature_1995,abrikosov_parity_1995,hague_d-wave_2006}.  For these cases, we can always return to Eq.(\ref{Eliashberg_v2}). 

\section{Implementation}
In this section, we discuss the implementation of the formalism developed in the last section using the PAW method.  An implementation for liquids using the same method is presented in Ref.~\onlinecite{chen_first-principles_2021}. Here, we discuss issues specific to an implementation for solids. 

\subsection{Green's function}\label{G_PAW}
The Green's function $\hat{\mathcal{G}}[\bm{R}(\tau)]$ is a central quantity in our formalism. In this subsection, we discuss its evaluation using the PAW method.


The PAW method is an efficient method for determining electron states in DFT calculations~\cite{blochl_projector_1994,kresse_ultrasoft_1999}.  In the method, calculations are performed in a space of smooth pseudo (PS) wave functions $\tilde{\psi}$. True all-electron (AE) wave functions $\psi$ are related with the PS wave functions by a linear transformation

\begin{eqnarray}\label{Trans}
	|\psi\rangle=\hat{T}(\bm{R})|\tilde{\psi}\rangle
\end{eqnarray}
with
\begin{eqnarray}\label{Trans_op}
	\hat{T}(\bm{R})=\hat{\mathbb{I}}
	+\sum_{ia}\left(|\phi_{i}^a(\bm{R}_a)\rangle-|\tilde{\phi}_{i}^a(\bm{R}_a)\rangle\right)\langle \tilde{p}_i^a(\bm{R}_a)|,\quad 
\end{eqnarray}
where  $|\phi_{i}^a(\bm{R}_a)\rangle$ and $|\tilde{\phi}_{i}^a(\bm{R}_a)\rangle$ denote the $i$-th AE and PS partial waves around ion $a$, respectively, and $\langle\tilde{p}_i^a(\bm{R}_a)|$ is the projector function which is bi-orthonormal to PS partial waves. We note that the transformation operator depends on the ion positions $\bm{R}\equiv\{\bm{R}_a\}$.




In Ref.~\onlinecite{chen_first-principles_2021}, it is shown that the Green's function, under the quasi-static approximation~\cite{liu_superconducting_2020}, can be determined by
\begin{eqnarray}\label{Green1}
	\hat{\mathcal{G}}(i\omega_j,\bm{R})=	\hat{T}(\bm{R})\hat{\tilde{\mathcal{G}}}(i\omega_j,\bm{R})\hat{T}^\dagger(\bm{R}),
\end{eqnarray}
and
\begin{eqnarray}\label{Green2}
  \hat{\tilde{\mathcal{G}}}(\mathrm{i}\omega_j,\bm{R})=\hbar\left[(\mathrm{i}\hbar\omega_j+\varepsilon_F)\hat{S}(\bm{R})-\hat{\tilde{H}}(\bm{R})\right]^{-1},
\end{eqnarray}
where $\hat{\tilde{H}}(\bm{R})$ is the Hamiltonian matrix in the PS space for a static ion configuration $\bm{R}$, and $\hat{S}(\bm{R})=\hat{T}^{\dagger}(\bm{R})\hat{T}(\bm{R})$ is the overlap matrix between PS waves.

For a solid, it is more convenient and efficient to work completely in a PS space. However, a PS space changes when ions move. To this end, we define a common PS space which is associated with the equilibrium ionic configuration $\bm{R}^{(0)}$.  We have the relation
\begin{multline}\label{Green_phi0}
  \left\langle\psi\left|\hat{\mathcal{G}}(i\omega_j,\bm{R})\right|\psi^{\prime}\right\rangle \\
	=
	\left\langle\tilde{\psi}\left|
	\hat{S}(\bm{R}^{(0)},\bm{R})
	\hat{\tilde{\mathcal{G}}}(\mathrm{i}\omega_{j}, \bm{R})
	\hat{S}(\bm{R},\bm{R}^{(0)})
	\right|\tilde{\psi}^{\prime}\right\rangle,\quad
 \end{multline}
where $\psi$ ($\tilde\psi$) and $\psi^{\prime}$ ($\tilde\psi^{\prime}$) are two states in the AE (common PS) space, and we define an overlap matrix:
\begin{eqnarray}\label{overlap}
	\hat{S}(\bm{R},\bm{R}^{(0)}) = \hat{T}^\dagger(\bm{R})\hat{T}(\bm{R}^{(0)}).
\end{eqnarray}

To evaluate the overlap matrix $\hat{S}(\bm{R},\bm{R}^{(0)})$, we substitute Eq.~(\ref{Trans_op}) in Eq.~(\ref{overlap}), and obtain
\begin{multline}\label{Gen_overlap}
	\hat{S}(\bm{R},\bm{R}^{(0)}) = \hat{T}^\dagger(\bm{R})+\hat{T}(\bm{R}^{(0)}) -\hat{\mathbb{I}}\\
	+ \sum_{i,j,a,b} \Big|\tilde{p}_i(\bm{R}_a)\Big\rangle
	Q_{ij}^{ab}(\bm{R}_b^{(0)}-\bm{R}_a)
	\Big\langle\tilde{p}_{j}(\bm{R}_b^{(0)})\Big|,
\end{multline}
where we define a set of coefficients
\begin{multline}
	Q_{ij}^{ab}(\bm{R}_b^{(0)}-\bm{R}_a)\\
	=\left\langle\phi_i^a(\bm{R}_a)-\tilde{\phi}_i^a(\bm{R}_a)\Big|\phi_{j}^b(\bm{R}_b^{(0)})-\tilde{\phi}_{j}^b(\bm{R}_b^{(0)})\right\rangle,
\end{multline}
which is the overlap between partial waves defined for two different centers. The coefficients can be evaluated numerically~(see Appendix~\ref{Qij}).  We note that for consistency and accuracy, we also need to use Eq.~(\ref{Gen_overlap}) to determine the regular overlap matrix $\hat{S}(\bm{R})\equiv\hat{S}(\bm{R},\bm{R})$~\cite{chen_first-principles_2021}.

We can use Eq.~(\ref{Green_phi0}) to simplify the determination of the generalized Bloch states. We expect that they are close to the regular Bloch states which are obtained by assuming ions fixed at equilibrium positions.  We thus first apply Eq.~(\ref{Green_phi0}) to determine the matrix elements of Green's function in the regular Bloch basis.  After the configuration average, we can determine the matrix of $H_{\mathrm{eff}}$ by applying Eq.~(\ref{Heff}), also in the regular Bloch basis.  We then diagonalize the matrix, and obtain generalized Bloch states as linear superpositions of the regular ones. In this way, all calculations can be performed in the common PS space.


\subsection{Discretization errors of the effective interaction}\label{os_tau}
\begin{figure}
	\includegraphics[width=8.6cm]{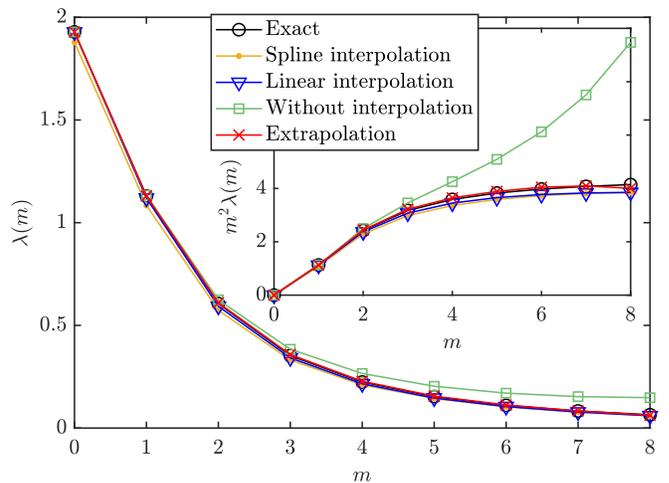}
	\caption{\label{fig:Di} $\lambda(m)$ obtained by applying the extrapolation approach as well as the oversampling approach with different interpolation methods: without oversampling, linear interpolation and spline interpolation. In the inset, we show the asymptotic behavior of $m^2\lambda(m)$. The exact result is also shown.  The results are for a model system with six harmonic modes of frequencies $\hbar\omega/k_B=\{3000\mathrm{K},2300\mathrm{K},2000\mathrm{K},1500\mathrm{K},1300\mathrm{K},500\mathrm{K}\}$ and $|g|^2=\{1.6,1.2,2.5,1.5,1,0.5\}$ at $T=190$~K. 
	Extrapolation is performed from values solved with five bead numbers $N_b=\{8,12,16,24,32\}$. Oversampling is performed from values solved with a single bead number $N_b=16$.}
\end{figure}

In PIMD simulations, one discretizes the time domain to a finite number of beads. The bead number is chosen by requiring that the maximum Matsubara frequency $\pi N_b k_B T$ is much larger than vibration frequencies, so that the imaginary-time evolution of ions can be correctly sampled. The finite number of beads introduce discretization errors when we perform Fourier transformations, making the determination of high-frequency components of effective interaction inaccurate. This is the case for the determination of the EPC parameters $\lambda(m)$, which are inputs of the linearized Eliashberg equation, and directly determine $T_{c}$.  In theory, one expects that $m^{2}\lambda(m)$ approaches to a constant when the Bosonic Matsubara frequency $\nu_{m} = 2\pi m/\hbar\beta$ is much larger than the Debye frequency of phonons.  The asymptotic behavior is useful for determining an average phonon frequency~\cite{allen_transition_1975}
\begin{eqnarray}\label{w2}
	\bar{\omega}_2=\lim_{m\rightarrow\infty} \frac{2\pi}{\hbar\beta} \sqrt{\frac{m^2\lambda(m)}{\lambda(0)}},
\end{eqnarray}
which also enters the Eliashberg equation by correcting the Morel Andersen pseudopotential $\mu^*$~\cite{morel_calculation_1962,allen_transition_1975}.  However, due to the discretization errors, such asymptotic behavior cannot be actually observed in PIMD simulations using a small number of beads $N_b$, as shown in Fig.~\ref{fig:Di}.

\subsubsection{Extrapolation to the quantum limit}

\begin{figure}
	\includegraphics[width=8.6cm]{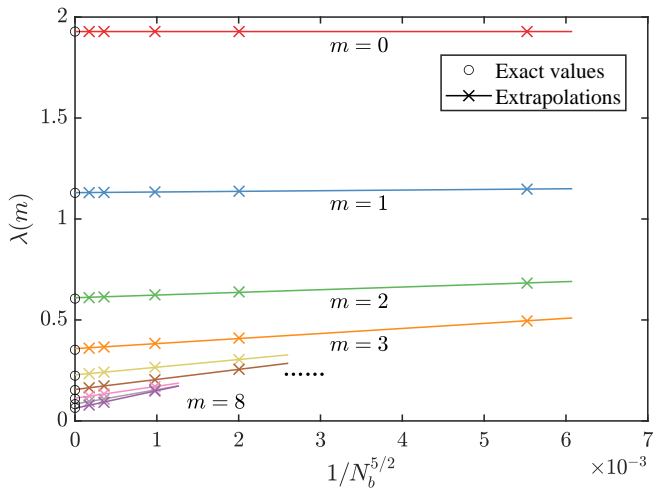}
	\caption{\label{fig:extra_har}
		Extrapolation of $\lambda(m)$ with respect to the bead number $N_b$. Exact results are shown for comparison. The results are for a harmonic model with The results are for a model system with six harmonic modes of frequencies $\hbar\omega/k_B=\{3000\mathrm{K},2300\mathrm{K},2000\mathrm{K},1500\mathrm{K},1300\mathrm{K},500\mathrm{K}\}$ and $|g|^2=\{1.6,1.2,2.5,1.5,1,0.5\}$ at $T=190$~K.  The model is solved with $N_b=\{8,12,16,24,32\}$.}
\end{figure}

One way to solve the issue is to perform an extrapolation to the infinite--$N_{b}$ limit. This is to view $\hat{W}(\nu_m)$ or $\lambda(m)$ as a function of the bead number $N_b$, and calculate the correct values of the quantities by extrapolating $N_b$ to infinity. 
We test the method for various model harmonic systems. Each model consists of several harmonic oscillators with different frequencies, and the effective interaction is determined by $W(\tau) = \sum_{l}|g_{l}|^{2} D_{l}(\tau)$, where $D_{l}(\tau)$ is the Green's function of the $l$-th harmonic mode, and $|g_{l}|^{2}$ denotes the EPC strength of the mode. The model can be solved using different bead numbers, and then be extrapolated.
In Fig.~\ref{fig:extra_har}, we show such an example. We find that $\lambda_{N_b}(m)$ is approximately proportional to $(N_b)^{-5/2}$.
The extrapolated values are shown in Fig.~\ref{fig:Di}. We see that the extrapolation converges well to the exact result.
   
However, the extrapolation method requires multiple PIMD simulations with different numbers of beads.  For our calculation based on the first-principles method, it is too expensive, and may not be feasible for complex materials with large unit cells (e.g. H$_{3}$S).  We thus seek for alternative methods which are less demanding in computational resources.

\subsubsection{Oversampling in the time domain}
An alternative method we develop is based on oversampling in the time domain.
We expect that the effective interaction $\hat{W}(\tau)$ is a smooth function of the imaginary time $\tau$.
While a PIMD simulations only determines the values of $\hat{W}(\tau)$ for a discrete set of the imaginary time \{$\tau_a=\hbar (a-1)/N_b k_B T$, $a=1,\cdots,N_b$\},  we can over-sample the original data to a much denser time-grid by employing an interpolation method. The over-sampled data is then Fourier transformed to obtain frequency components of the effective interaction and $\lambda(m)$. We find that resulting $\lambda(m)$ does have the correct asymptotic behavior, and converges reasonably well to the exact result, as shown in Fig.~\ref{fig:Di}. 

%

An obvious freedom of the approach is the choice of the interpolation method. To see which method is optimal for our purpose, we again test our oversampling procedure for model harmonic systems. Effective interactions of the models are oversampled, and compared with the exact result. In Fig.~\ref{fig:Di}, $\lambda(m)$ obtained from the oversampling approach using different interpolation methods are shown.  We find that all interpolation methods yield the correct asymptotic behavior in large $m$, and the linear interpolation gives $\lambda(m)$ closest to their exact values. Based on the observation, we choose to use the linear interpolation method in this study.

The extrapolation and oversampling  methods developed above are based on harmonic models. In practice, we find that these methods also apply well for anharmonic solids investigated in this study.  This is tested in Appendix~\ref{os_test} by applying both the extrapolation and the oversampling approach to calculate $\lambda(m)$ 
in the anharmonic solid H$_3$S. We find both the approaches yield the correct asymptotic behavior of $\lambda(m)$.  Results obtained by the two different approaches coincide well. Considering that the oversampling approach is much less demanding for computational resources, we choose to use the approach in our study.

%

\subsection{Temperature dependence of EPC parameters}\label{Td}
To determine $T_c$, one needs to find the temperature at which a non-negative eigenvalue $\rho$ of the Eliashberg equation~(\ref{Eliashberg_v2}) first appears. It usually requires at least two PIMD simulations at different temperatures, ideally one above $T_{c}$ and another below. This is how $T_{c}$ is estimated in a liquid, since the effective pairing interaction in a liquid may strongly depend on the temperature.  On the other hand, for solids, the temperature dependence of the effective interaction is usually weak.  It is then possible to calculate at only one temperature close to $T_{c}$.  

We basically assume that vibrational properties remain unchanged within a small temperature range.  Therefore, the effective pairing interaction is nearly temperature independent and close to the one we calculate.  In this case, EPC parameters for different temperatures can be determined by a single frequency-dependent function $\Lambda(\nu)$:
\begin{equation}\label{lambda_T}
	\lambda(m,T) = \Lambda\bm{(}\nu_m(T)\bm{)},
\end{equation}
where $\nu_{m}(T)\equiv2m\pi k_BT/\hbar$ is the Bosonic Matsubara frequency at temperature $T$. By calculating $\lambda(m,T_0)$ at a given temperature $T_0$, we can construct an interpolation formula for $\Lambda(\nu)$. $\lambda(m,T)$ at other temperatures can then be inferred from Eq.~(\ref{lambda_T}).  

We test the scheme in Appendix~\ref{lambda_T_test}, and find the scheme also work well for anharmonic solids like H$_3$S.

\section{Metallic deuterium}\label{Hydrogen}
Metallic hydrogen is long believed to be a candidate of high-$T_c$ superconductors because of its high vibrational frequencies and strong EPC~\cite{ashcroft_metallic_1968,mcmahon_high-temperature_2011,borinaga_anharmonic_2016}. Based on \textit{ab initio} calculations, it is predicted that hydrogen atoms form an atomic metal under a pressure above $500$~GPa~\cite{mcmahon_ground-state_2011,azadi_dissociation_2014} and have a $T_c$ over 300~K~\cite{mcmahon_high-temperature_2011}.  While anharmonicity in the system is shown to be weak~\cite{borinaga_anharmonic_2016}, PIMD simulations indicate that quantum fluctuation due to tunnelings of hydrogen atoms are strong~\cite{chen_quantum_2013}. As a result, at $T_{c}$, metallic hydrogen could be a liquid instead of a solid~\cite{liu_superconducting_2020,chen_first-principles_2021}. Here, we choose to study metallic deuterium, which is still a solid at $T_{c}$ because quantum fluctuations are suppressed by the heavier mass of deuterium atoms~\cite{liu_superconducting_2020}.  We expect that metallic deuterium has anharmonicity further suppressed compared to metallic hydrogen, and can be viewed approximately as a harmonic solid.  We apply SPIA to it. Results are compared with those obtained from a standard harmonic approach.  This serves as a benchmark test for the accuracy of our approach and implementation.

\subsection{Numerical details}
We calculate deuterium with the structure of $I4_1/amd$ space group at $500$~GPa.  Structure parameters of metallic hydrogen are used~\cite{mcmahon_high-temperature_2011}.  PIMD simulations are performed in a Born-von Karman diagonal $6\times6\times6$ supercell, at $250$~K, using the CPU and GPU version of the Vienna ab initio Simulation Package (VASP) code~\cite{kresse_efficient_1996,kresse_ultrasoft_1999}. The PAW method is used to describe the ion-electron interaction, and the Perdew-Burke-Ernzerhof (PBE) functional~\cite{perdew_generalized_1996} is used to describe the exchange-correlation effect. An energy cutoff of $450$~eV for plane waves is used to expand electron wave functions.  A $3\times3\times3$ $\Gamma$-centered k-point mesh is used to sample the Brillouin zone for the supercell. The Andersen thermostat~\cite{andersen_molecular_1980} is used to control the temperature of the canonical (NVT) ensemble, and ion velocities are randomized according to the Maxwellian distribution every 25~fs. A time step of $0.5$~fs and an overall simulation time of $3$~ps with bead number $N_b=24$ is used to simulate the quantum system. We work under the normal-mode representation to describe the inter-bead oscillations~\cite{ceriotti_efficient_2010}. Artificial masses are set for different normal modes, so that they can oscillate on the same time scale as phonons~\cite{li_computer_2018}.

Sampled ionic configurations are analyzed using our implementation in MATLAB. The program can be found in Ref.~\footnote{The source codes of the program can be downloaded from \url{https://github.com/Haoran-Chen-1115/SPIA}.} .
In the analysis, the Brillouin zone for the supercell is sampled using a denser $6\times6\times6$ k-mesh, which is equivalent to a $36\times36\times36$ mesh of the primitive Brillouin zone. Hamiltonians at these k points are reconstructed by using outputs (e.~g., the local density and pseudo-potential) of VASP in the PIMD simulations. Green's functions are determined by using Eq.~(\ref{Green1}) with
$j=16$.  An irreducible wedge of the k-mesh and a $0.03$-Ry Gaussian smearing are then used for performing the summation in Eq.~(\ref{lambda_avg}), for calculating EPC parameters $\lambda(m)$.

Harmonic calculations are performed with DFPT~\cite{baroni_phonons_2001} implemented in the QUANTUM ESPRESSO package~\cite{giannozzi_quantum_2009}, with the ion-electron interaction described using ultrasoft pseudopotential generated by A. Dal Corso using the code Standard solid-state pseudopotentials (SSSP)~\cite{prandini_precision_2018}.
An energy cutoff of $80$~Ry is used to expand the wave functions. A $36\times36\times36$ k-mesh is used for calculating phonon frequencies. EPC matrix elements on the $6\times6\times6$ $\bm{q}$-grid are used to calculate EPC parameters,  a $0.03$-Ry Gaussian smearing is used for the Fermi-surface summation. These parameters are chosen to be consistent with their counterparts in SPIA calculations, for the convenience of a comparison. For the same reason, we use the regular Bloch states in SPIA calculations, since corrections due to the generalized Bloch states are expected to be weak, and the harmonic approach always uses the regular ones.

\subsection{Results}\label{Results_D}

\begin{figure}[t]
	\includegraphics[width=8.6cm]{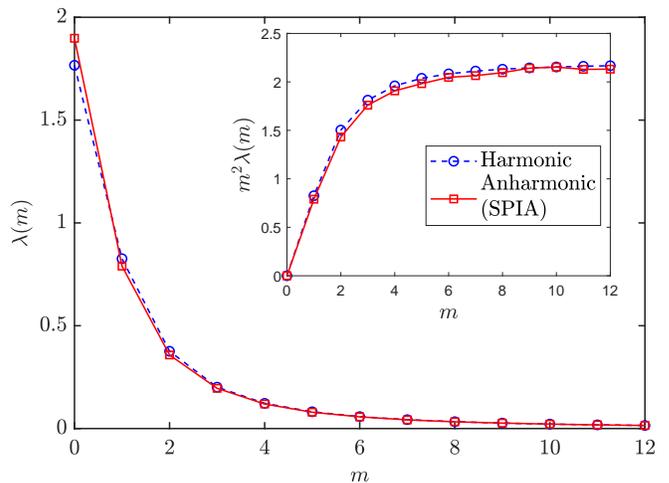}
	\caption{\label{fig:H_lambda} EPC parameters $\lambda(m)$ of metallic deuterium under $500$~GPa calculated by using the standard harmonic approach and SPIA.  The asymptotic behavior of $m^2\lambda(m)$ is also shown in the inset.}
\end{figure}

\begin{table}
	\begin{ruledtabular}
		\begin{tabular}{llccccc}
			System & Method
			& $\lambda(0)$  &  $\lambda(1)$  &  $\lambda(2)$  & $\bar{\omega}_2$    & $T_c$   \\
			\midrule
			\multirow{2}{*}{\shortstack{$D$\\$(I4_1/amd)$}}
			& DFPT
			&  1.767        &  0.826         &  0.376         & 1752              & 226     \\
			& SPIA
			&  1.898        &  0.790         &  0.358         & 1657              & 213     \\
			\midrule
			\multirow{2}{*}{\shortstack{$T$\\$(I4_1/amd)$}}
			& DFPT
			&  1.767        &  0.957         &  0.480         & 1431              & 185     \\
			& SPIA
			&  1.820        &  0.918         &  0.458         & 1392              & 177     \\
		\end{tabular}
	\end{ruledtabular}
	\caption{\label{tab:result_H}%
		First few EPC parameters at 250~K for deuterium and at 170~K for tritium, average phonon frequency $\bar{\omega}_2$ (in K), and predicted $T_{c}$ (in K) of metallic deuterium at $500$~GPa. 
		We set $\mu^{*}=0.089$. 
	}
\end{table}


The EPC parameters $\lambda(m)$ are calculated using both the SPIA and the harmonic approach, and shown in Fig.~\ref{fig:H_lambda}. The values of the first few EPC parameters as well as $\bar{\omega}_{2}$ are shown in Table~\ref{tab:result_H}.  It is evident that the two calculations coincide rather well.  
To explain the small quantitative differences ($\sim 7\%$), we also calculate the parameters for tritium, which has heavier mass. We find that the differences are further reduced (to $\sim 3\%$). This suggests the small differences could be due to small residual effects of anharmonicity, or vertex corrections ($\propto \sqrt{m_{e}/M_{\mathrm{ion}}}\sim 1.6\%$ for deuterium and $\sim1.3\%$ for tritium) ignored in the harmonic calculation.

\begin{figure}
	\includegraphics[width=8.6cm]{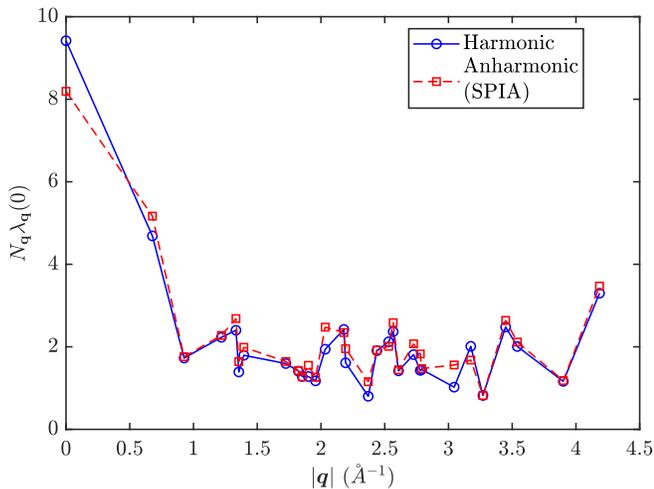}
	\caption{\label{fig:D_lamq} $N_{\bm{q}}\lambda_{\bm{q}}(0)$ for solid deuterium as a function of $|\bm{q}|$. Results from harmonic and SPIA are shown. $N_{\bm{q}}$ is the total number of $\bm{q}$-points sampled in the simulation.
	}
\end{figure}

The SPIA and harmonic calculations actually coincide well to details.  To see this, we define the $\bm{q}$-resolved EPC parameters $\lambda_{\bm{q}}$:
\begin{multline}\label{lambda_q}
	\lambda_{\bm{q}}(m)=-\frac{1}{N(\varepsilon_F)}\sum_{nn',\bm{k}}
	W_{n\bm{k},n'\bm{k}-\bm{q}}(m)\\
	\times\delta(\varepsilon_{n\bm{k}}-\varepsilon_F)\delta(\varepsilon_{n'\bm{k}-\bm{q}}-\varepsilon_F),
\end{multline}
where $\bm{q}$ is the wave-vector of phonons mediating the effective interaction. In Fig.~\ref{fig:D_lamq}, we show $\lambda_{\bm{q}}(0)$ as a function of $|\bm{q}|$. It is evident that the two results are close even they are obtained from two completely different approaches. It indicates that our approach can well reproduce results from a standard harmonic approach for (nearly) harmonic solids.  The observation also provides the confidence that the SPIA approach is properly implemented.

With the EPC parameters, we can estimate $T_c$ by solving the linearized Eliashberg equation~(\ref{Eliashberg_v2}). We use a Morel-Andersen pseudopotential of $\mu^{*}=0.089$~\cite{mcmahon_high-temperature_2011}, and obtain $T_c=226$~K and $213$~K from the harmonic approach and the SPIA, respectively.
The suppression of the predicted $T_c$ after considering anharmonic effects in SPIA is small ($\sim 6\%$).  It is consistent with previous studies based on SSCHA~\cite{borinaga_anharmonic_2016}, which shows that anharmonicity has small effects on the superconductivity of metallic hydrogen.

\section{Hydrogen sulfide}\label{H3S}
Hydrogen sulfide is the first high-$T_c$ conventional superconductor ever discovered~\cite{drozdov_conventional_2015}.  It was first predicted in theory by using the standard harmonic approach~\cite{duan_pressure-induced_2014}.  The predicted $T_c$ from the strong-coupling Eliashberg theory is about $60$~K higher than that observed in experiments~\cite{errea_high-pressure_2015}. The deviation is ascribed to anharmonicity and quantum effects of hydrogen atoms~\cite{errea_high-pressure_2015}. In this section, we apply the SPIA to hydrogen sulfide and compare it with the harmonic approach and the SSCHA approach~\cite{errea_high-pressure_2015}.

\subsection{Numerical details}
The structure parameters of H$_3$S are adopted from Ref.~\onlinecite{duan_pressure-induced_2014}. Electronic structure is described in the same way as in metallic deuterium. An energy cutoff of $600$-eV for the plane waves is used to expand electron wave functions. PIMD simulations are performed at $190$~K under pressure of 200 GPa, and in non-diagonal supercells (see the next subsection).  The temperature is controlled with Andersen thermostat, and ion velocities are randomized according to the Maxwellian distribution every 75~fs. A time step of $1.5$~fs and a time length no less than $8.25$~ps with bead number $N_b=16$ is used to simulate the quantum system.  
$k$-grids equivalent to the $12\times12\times12$ grid of the primitive cell is used to sample the Brillouin zone of supercells.  In the follow-up analysis, a denser $24\times24\times24$ $k$-grid of the primitive cell is used to calculate the EPC parameters $\lambda(m)$, for which a smaller energy cutoff of $450$-eV is used. Other parameters are the same as those used in the simulations of deuterium. 

Harmonic calculations are also performed in the same way as in solid deuterium. The ion-electron interaction are described using ultrasoft pseudopotential generated by A. Dal Corso using the code SSSP. An energy cutoff of $80$~Ry is used to expand the wave functions. A $24\times24\times24$ k-mesh is used for calculating phonon frequencies. EPC matrix elements on the $6\times6\times6$ $\bm{q}$-grid are used to calculate EPC parameters, and a $0.02$-Ry Gaussian smearing is used for Fermi-surface summation. 

\subsection{Non-diagonal supercells for sampling \textbf{q}  points}
In the simulation of hydrogen sulfide, we apply the non-diagonal-supercell technique to improve the efficiency of sampling the quasi-wave-vector $\bm{q}$ of phonons~\cite{lloyd-williams_lattice_2015}. With the technique, all $\bm{q}$ points on a uniform $N\times N\times N$ grid can be sampled by using a number of supercells containing at most $N$ primitive cells. PIMD simulations can then be preformed in smaller supercells, and computational cost is greatly reduced.

To apply the technique for the PIMD simulations, 
we treat a regular $3\times3\times3$ supercell as the ``primitive cell'' for constructing non-diagonal supercells. The PIMD simulations are then preformed in supercells containing two $3\times3\times3$ ``primitive cells''. In this way, we can sample all points in the irreducible part of a $6\times6\times6$ $\bm{q}$-grid by simulating in two non-diagonal supercells. 

We test the technique in Appendix~\ref{NDcell_t}.  We find that smaller non-diagonal supercells can indeed well reproduce results of a full supercell.

\begin{figure}[t]
	\includegraphics[width=8.6cm]{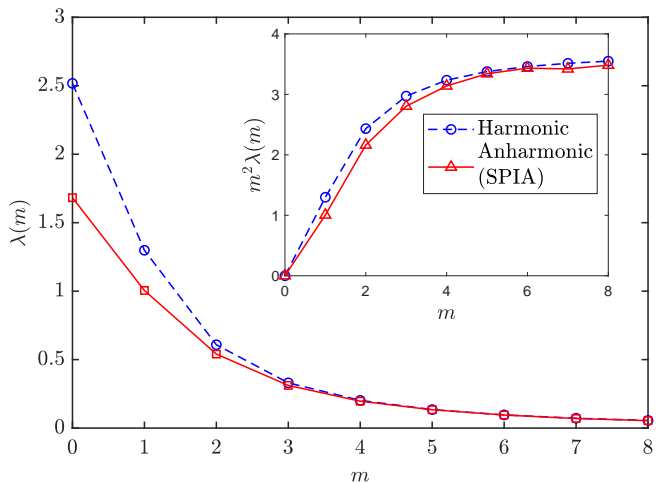}
	\caption{\label{fig:H3S_lambda} EPC parameters $\lambda(m)$ of H$_3$S determined by the harmonic approach and SPIA.  The inset shows the asymptotic behavior of $m^2\lambda(m)$.}
\end{figure}

\begin{table}
	\begin{ruledtabular}
		\begin{tabular}{llccc}
			System & Method
			& $\lambda(0)$     & $\bar{\omega}_2$     & $T_c$   \\
			\midrule
			\multirow{4}{*}{\shortstack{H$_3$S\\$(Im\bar{3}m)$}}
			& DFPT
			&  2.517        &  1441              & 243	     \\
			& SSCHA
			&  1.840        &     --               & 215\footnote{Result with $\mu^{*}=0.12$ provided in the supplementary of Ref.~\onlinecite{errea_high-pressure_2015}. A larger $\mu^{*}=0.16$ yields $T_c=$194~$K$.}     \\
			& SPIA
			&  1.682        & 1735               & 201     \\
			& SPIA-0
			&  1.605        & 1734               & 192     \\
			\cmidrule{2-5}  
			& Experiment
			&     --        &     --               & 190       
		\end{tabular}
	\end{ruledtabular}
	\caption{\label{tab:result_H3S}%
		Mass enhancement factor $\lambda(0)$,  average phonon frequency $\bar{\omega}_2$ (in K), and $T_{c}$ (in K) for H$_3$S of $Im\bar{3}m$ space group at $200$~GPa. SPIA-0 indicates using the regular Bloch basis instead of the generalized one. 
		 $\mu^{*}=0.12$ is used when determining $T_c$.}
\end{table}

\subsection{Results}
We determine the EPC parameters of H$_3$S, shown in Fig.~\ref{fig:H3S_lambda}.
Different from metallic deuterium, anharmonic effects play important roles in H$_3$S. We see that the EPC parameters determined by SPIA are strongly suppressed from those by the harmonic approach, and $m^{2}\lambda(m)$ approaches to a lower asymptotic value at large $m$. It results in an enhanced average phonon frequency $\bar{\omega}_2$, shown in Table~\ref{tab:result_H3S}.  The big differences indicate that ion vibrations are far from being harmonic. Setting $\mu^{*}=0.12$~\cite{sano_effect_2016}, we find that the predicted $T_c$ is suppressed from $243$~K of the harmonic calculation to $201$~K of SPIA, a $17\%$ supression. The result is close to $190$~K observed in experiments. Compared to the SSCHA prediction $T_c=215$~K (for $\mu^{\ast}=0.12$)~\cite{errea_high-pressure_2015}, our prediction is closer to the experimentally observed $T_{c}$, provided a reasonble value of $\mu^{\ast}$ is used.

\begin{figure}[t]
	\includegraphics[width=8.6cm]{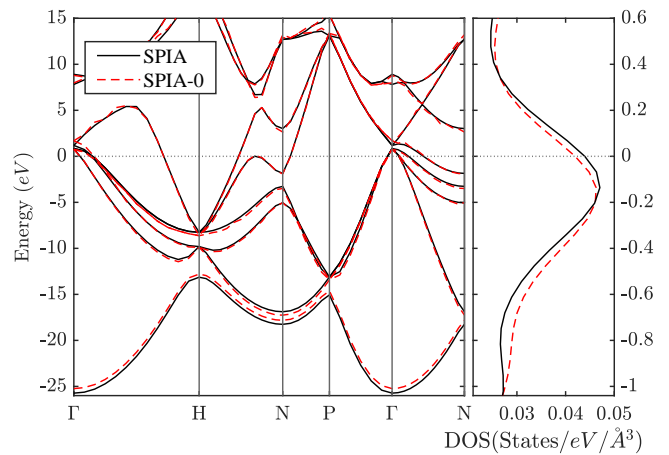}
	\caption{\label{fig:H3S_band} Left panel: band dispersion of H$_3$S without and with the band renormalization. Right panel:  DOS near the Fermi level.}
\end{figure}

To see the effect of band renormalization in the generalized Bloch basis, we also preform the calculation using the regular Bloch basis.
The result is also shown in Table~\ref{tab:result_H3S}, denote as SPIA-0.
Band structures and density of states (DOS) with and without the renormalization are shown in Fig.~\ref{fig:H3S_band}. We find that the effect of the renormalization is small for electronic structure near the Fermi surface.  It leads to an enhancement of DOS at the Fermi level $N(\varepsilon_F)$ by $5\%$. Accordingly, $\lambda(0)$ is enhanced by the same ratio. It enhances the predicted $T_c$ from $192$~K of SPIA-0 to $201$~K of SPIA. The observation is consistent with the prediction of Ref.~\onlinecite{sano_effect_2016}, which shows that the band renormalization would enhance the predicted $T_{c}$.  


\section{Summary}\label{Summary}
In summary, we develop a stochastic path integral approach for treating effects of anharmonicity and quantum fluctuations on superconductivity in anharmonic solids.  The formalism of SPIA is extended for solids by defining generalized Bloch basis. We implement the approach in DFT calculations using the PAW method. Issues associated with the implementation are solved. We find that our approach can well reproduce results of a harmonic solid (metallic deuterium), and predict $T_{c}$ closer to the experimentally observed one than the SSCHA approach for an anharmonic solid (hydrogen sulfide).

\begin{acknowledgments}
	We gratefully acknowledge discussions with Xin-Zheng Li and Xiao-Wei Zhang on the numerical details of PIMD simulations.
	The authors are supported by the National Key R\&D Program of China under Grand Nos.~2018YFA0305603 and 2021YFA1401900, the National Science Foundation of China under Grant No.~12174005.  The computational resources were provided by the High-performance Computing Platform of Peking University.
\end{acknowledgments}

\appendix
\section{Overlap between partial waves}\label{Qij}
In this appendix, we discuss the evaluation of the overlap $Q_{ij}^{ab}(\Delta\bm{R})$ introduced in Sec.~\ref{G_PAW}.  To reduce computational cost, we determine the overlap by interpolation, in combination with a rotational transformation. The details are as follows.

As a first step, we arrange the two centers along y axis. We then calculate the overlaps for a list of displacement vectors $\bm{d}\equiv\Delta\bm{R}=d\hat{\bm{y}}$ along the $y$-axis:
\begin{multline}
	Q_{ij}^{ab}(\bm{d})
	= \int d^3r \left(\phi_i^a(r)-\tilde{\phi}_i^a(r)\right)
	Y_{l_i,m_i}(\hat{\bm{r}})\\
	 \times
	\left(\phi_j^b(|\bm{r}-\bm{d}|)-\tilde{\phi}_j^b(|\bm{r}-\bm{d}|)\right)
	Y_{l_j,m_j}(\widehat{\bm{r}-\bm{d}}),
\end{multline}
where $Y_{lm}$ is the spherical harmonics of the real form.
Using the list of the overlap values, we can set up an interpolation function for determining overlaps of arbitrary distances.

For two centers with the displacement vector not along the $y$-axis,  we can determine the overlap by applying a rotational transformation. To do that, we make use of the transformation of a spherical harmonics under a rotation:
\begin{equation}
  Y_{lm}(\theta,\phi) = \sum_{m^{\prime}} R^{l}_{mm^{\prime}} Y_{lm^{\prime}}(\theta^{\prime},\phi^{\prime}).
\end{equation}
We can choose a $(\theta^{\prime}, \phi^{\prime})$ frame which transforms the displacement vector between the two centers in the $(\theta,\phi)$ frame
\begin{eqnarray}
	\bm{d} = \left( d \cos{\alpha}\sin{\beta}, d \sin{\alpha}\sin{\beta}, d \cos{\beta}\right)
\end{eqnarray}
to a vector along the $y$-axis. The corresponding transformation matrix $[R^{l}_{mm^{\prime}}]$ for $l=1$,  $m,m^{\prime}={-1,0,1}$ is
\begin{eqnarray}
	&&\hat{{R}}^{l=1}=\begin{bmatrix}
		\sin{\alpha}\sin{\beta} & -\sin{\alpha}\cos{\beta} & -\cos{\alpha}\\
		\cos{\beta} & \sin{\beta} & 0\\
		\cos{\alpha}\sin{\beta} & -\cos{\alpha}\cos{\beta} & \sin{\alpha}
	\end{bmatrix}.
\end{eqnarray}
Transformation matrices for $l>1$ can be obtained by applying a recursion relation~\cite{ivanic_rotation_1996}.

The overlap for an arbitrary displacement vector $\bm{d}$ can then be obtained by applying the transformation
\begin{eqnarray}
	Q_{ij}^{ab}(\bm{d})
	=\sum_{kl} \mathbb{R}_{ik} Q_{kl}^{ab}(d\hat{\bm{y}}) \mathbb{R}_{jl},
\end{eqnarray}
with $\mathbb{R}_{ij}\equiv{R}^{l_{i}}_{m_i m_j}\delta_{l_i,l_j}$.

\section{Extrapolation and oversampling of effective interactions in H$_3$S}\label{os_test}
In this Appendix, we test the extrapolation method and the oversampling method developed in Sec.~\ref{os_tau} for the anharmonic solid H$_3$S. We perform simulations in a $3\times3\times3$ supercell at $190 K$, with bead numbers $N_b=8,16,24$ and $32$, and calculate $\lambda_{N_b}(m)$ on a $8\times8\times8$ $k$-grid of the supercell. In all simulations, a time length of $3.5$ ps with a $1.0$ fs time step is used.

\begin{figure}
	\includegraphics[width=8.6cm]{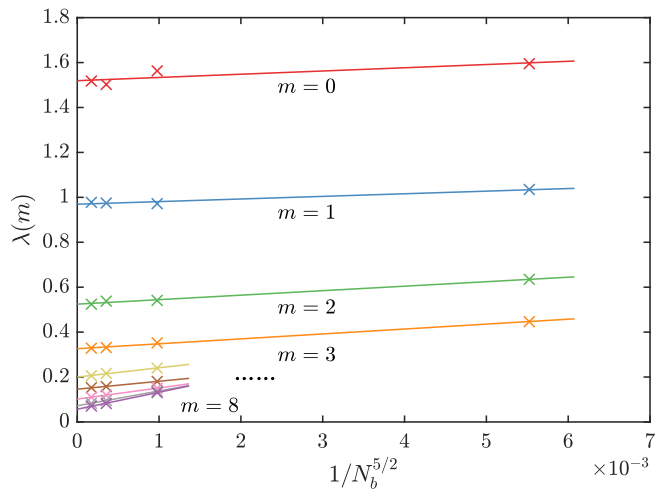}
	\caption{\label{fig:extra_H3S}
		Extrapolation of $\lambda(m)$ with respect to the bead number $N_b$. The results are for H$_3$S at $T=190$~K.}
\end{figure}

\begin{figure}
	\includegraphics[width=8.6cm]{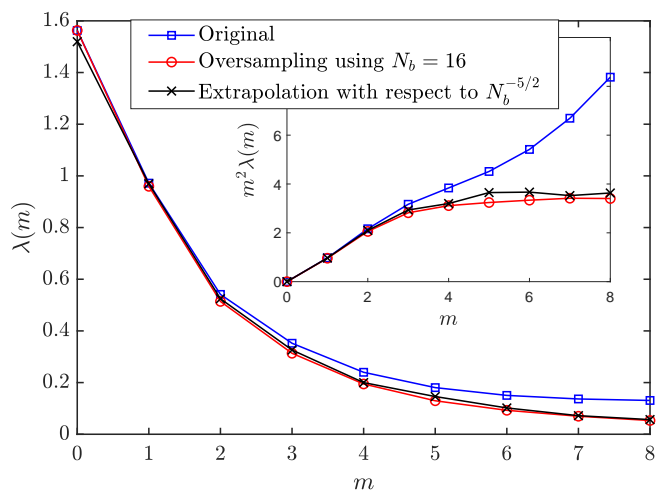}
	\caption{\label{fig:extra_os}
		EPC parameters $\lambda(m)$ of H$_3$S determined by using the oversampling approach and by extrapolating them to the quantum limit.  The inset shows the asymptotic behavior of $m^2\lambda(m)$.}
\end{figure}

First, we apply the extrapolation method. Results are shown in Fig.~\ref{fig:extra_H3S}.
Similarly to the harmonic case, we find that $\lambda_{N_b}(m)$ is approximately proportional to $(1/N_b)^{5/2}$. Slight deviations are due to statistical fluctuations.
We also apply the oversampling approach using $N_b=16$.
The results are compared with the extrapolated results in Fig.~\ref{fig:extra_os}. We find that both the approaches yield the correct asymptotic behavior of $\lambda(m)$, and the results coincide well. 

\section{Temperature dependence of EPC parameters}\label{lambda_T_test}
In this Appendix, we test the validity of Eq.~(\ref{lambda_T}) for H$_{3}$S, which is strongly anharmonic.  We therefore perform two independent PIMD simulations at $160$~K and $190$~K in a $3\times3\times3$ supercell, and calculate $\lambda(m)$ on a $10\times10\times10$ $k$-grid of the supercell.  Note that the supercell is smaller than the non-diagonal supercells we use in the main text.

\begin{figure}
	\includegraphics[width=8.6cm]{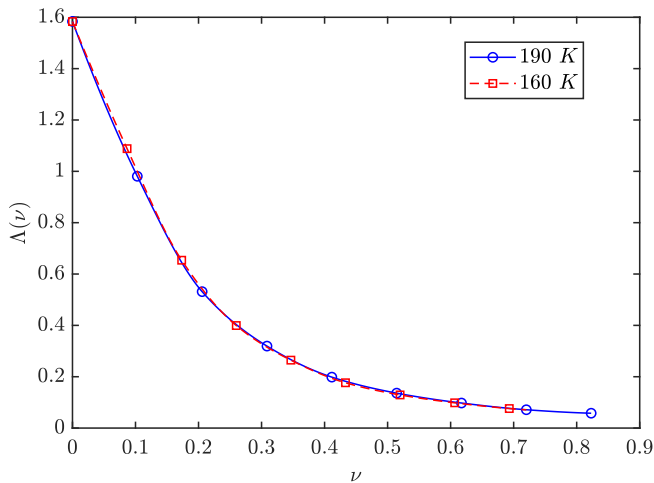}
	\caption{\label{fig:T_dep} $\lambda(m,T)$ of H$_{3}$S at $T=190$~K and 160~K (points), and corresponding interpolation functions $\Lambda(\nu)$ constructed from the $\lambda(m,T)$ data (lines).}
\end{figure}

In Fig.~\ref{fig:T_dep}, we show two interpolation functions $\Lambda(\nu)$ constructed from $\lambda(m, T)$ data of $160$~K and $190$~K, respectively.  It is evident that the two interpolation functions coincide well, and all points of $\lambda(m, T)$ from the two simulations at different temperatures fall onto a single curve.  $\bar{\omega}_2$ and $T_c$ from the two simulations are also in good agreement, as shown in Table~\ref{tab:T_dep}. It indicates that Eq.~(\ref{lambda_T}) is still valid for  H$_{3}$S, even though it is strongly anharmonic.

\begin{table}
	\begin{ruledtabular}
		\begin{tabular}{lcccc}
			System & $T_{\text{sim}}$
			& $\lambda(0)$   & $\bar{\omega}_2$    & $T_c$   \\
			\midrule
			\multirow{2}{*}{\shortstack{H$_3$S\\$(Im\bar{3}m)$}}
			& 190~$K$
			&  1.584         & 1847.0              & 200     \\
			& 160~$K$
			&  1.583         & 1819.6              & 201     \\
		\end{tabular}
	\end{ruledtabular}
	\caption{\label{tab:T_dep}%
		Mass enhancement factor $\lambda(0)$,  average phonon frequency $\bar{\omega}_2$ (in $K$), and predicted $T_{c}$ for H$_3$S, determined by simulations in a $3\times 3\times 3$ supercell and at temperatures $190$~K and $160$~K.}
\end{table}

\section{Test for non-diagonal supercell technique}\label{NDcell_t}
In this Appendix, we test the non-diagonal supercell method employed in the calculation of H$_{3}$S.  $\bm{q}$ points on a $4\times4\times4$ grid are sampled by directly using a $4\times4\times4$ diagonal supercell, and by using non-diagonal supercells built from $2\times2\times2$ diagonal ones.  In the test, PIMD simulations are performed using a time step of $1.5$~fs and a time length of $5.25$~ps with bead number $N_b=16$.

\begin{table}[tb]
	\begin{ruledtabular}
		\begin{tabular}{clccc}
			& $\bm{q}$ & ND1\footnote{Nondiagonal supercell with $\hat{\mathbb{S}}_1=\{\{4, 2, 2\},\{0, 2, 0\},\{0, 0, 2\}\}$.}
			& ND2\footnote{Nondiagonal supercell with $\hat{\mathbb{S}}_2=\{\{0, 2, 2\},\{2, 0, 2\},\{2, 2, 0\}\}$.} 
			& Diagonal\footnote{Diagonal supercell with $\hat{\mathbb{S}}=\mathrm{diag}\{4, 4, 4\}$.} \\
			\hline
			\multirow{8}{*}{$\lambda_{\bm{q}}(0)$}
			& ( 0 , 0 , 0 ) & 0.0094 & 0.0112 & 0.0082 \\
			& ( 0.25 , 0 , 0 ) & 0.0253 & -- &  0.0281 \\
			& ( 0.5 , 0 , 0 ) & 0.0162 & 0.0168 & 0.0171 \\
			& ( 0.25 , 0.25 , 0 ) & 0.0229 & -- & 0.0234 \\
			& ( 0.25 , 0.25 , 0.25 ) & -- & 0.0222 & 0.0227 \\
			& (-0.25 , 0.25 , 0.25 ) & -- & 0.0180 & 0.0177 \\
			& ( 0.5 , 0.5 , 0.25 ) & 0.0241 & -- & 0.0237 \\
			& ( 0.5 , 0.5 , 0.5 ) & 0.0142 & 0.0135 & 0.0128 \\
			\midrule
			$\lambda(0)$ & & \multicolumn{2}{c}{1.417} &  1.458\\
			$T_c$ & & \multicolumn{2}{c}{173} & 175 \\
		\end{tabular}
	\end{ruledtabular}
	\caption{\label{tab:ND_test1}%
		$\lambda_{\bm{q}}(0)$ at different $\bm{q}$ points, EPC parameter $\lambda(0)$ and predicted $T_c$ (in K) for non-diagonal and diagonal supercells. The shape of a non-diagonal supercell is specified by a transformation matrix $\hat{\mathbb{S}}$ (see Ref.~\onlinecite{lloyd-williams_lattice_2015}), which is indicated in the footnote.}
\end{table}

To test the method, we compare $\lambda_{\bm{q}}(0)$ at different $\bm{q}$ points. From Table~\ref{tab:ND_test1}, we see that the differences between different methods are small. They can be ascribed to finite size effect since non-diagonal supercells built from  $2\times2\times2$ cells are relatively small. We expect that for larger supercells, the convergence should be even better. Actually, in the main text, we perform simulations in non-diagonal supercells built from larger $3\times3\times3$ diagonal cells. We find $\lambda_{\bm{q}}(0)$ for $\bm{q}=0$ obtained from different non-diagonal supercells are almost identical. 

\section{Tests of convergence}\label{conv_test}
In this Appendix, we test the convergence of our calculations with respect to the bead number. 

\begin{figure}
	\includegraphics[width=8.6cm]{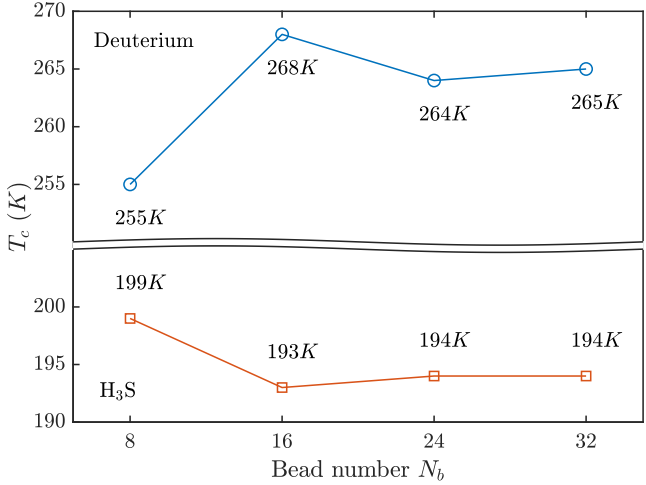}
	\caption{\label{fig:conv}
		The convergence test of $T_c$ with respect to the number of beads in deuterium (top) and hydrogen sulfide (bottom). The values are calculated using $\mu^*=0.089$ in deuterium and $\mu^*=0.12$ in H$_3$S. Simulations are performed in $3\times3\times3$ supercells in both systems. }
\end{figure}

\begin{figure}
	\includegraphics[width=8.6cm]{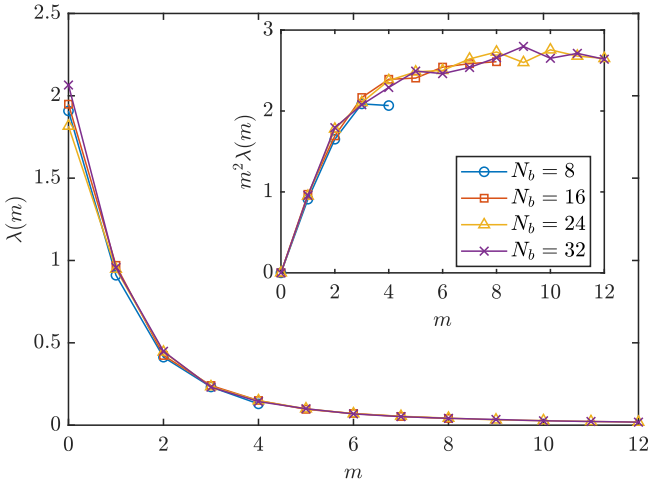}
	\caption{\label{fig:conv_D}
		The convergence test of EPC parameters $\lambda(m)$ with respect to the number of beads in Deuterium.}
\end{figure}

\begin{figure}
	\includegraphics[width=8.6cm]{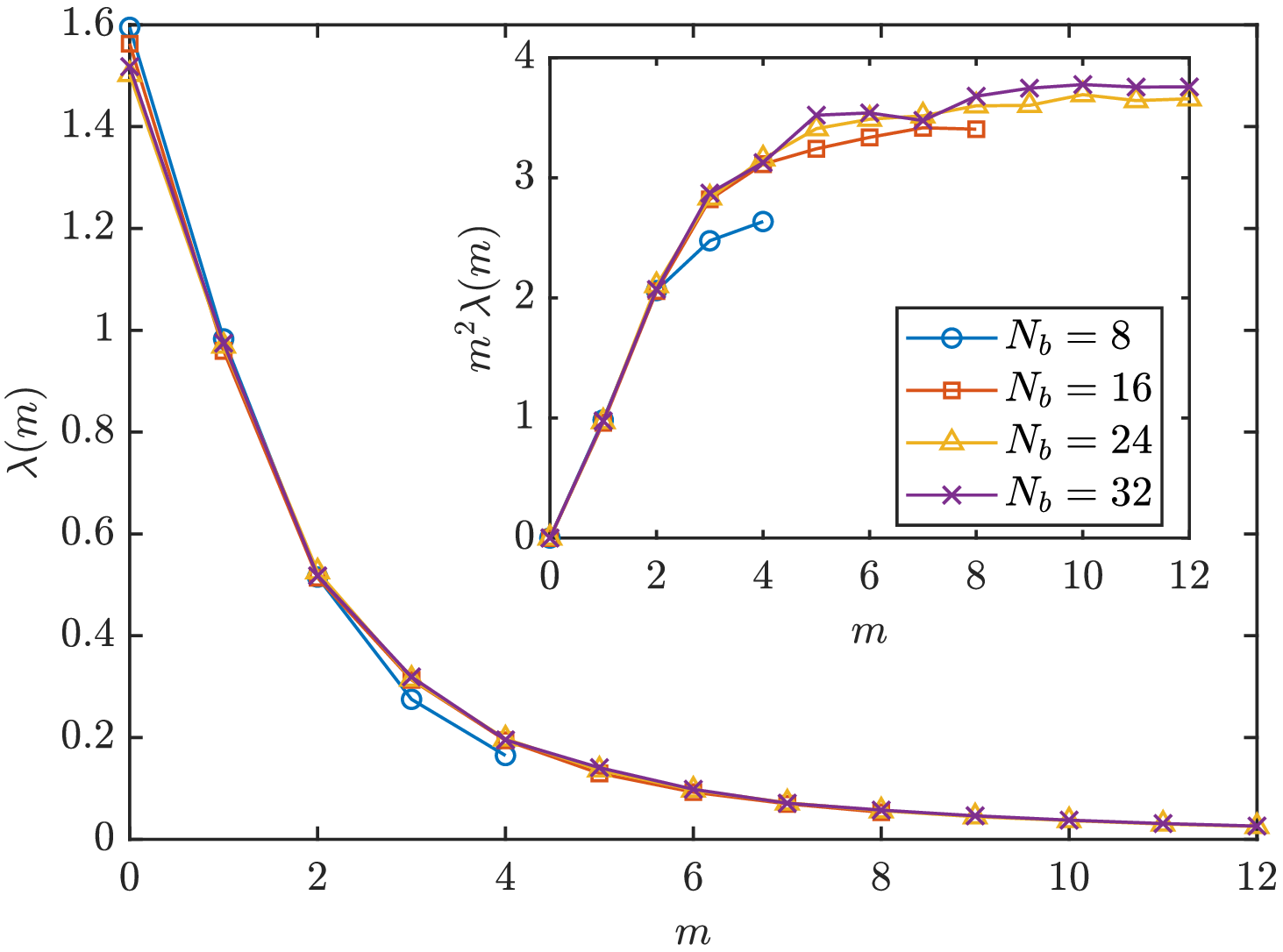}
	\caption{\label{fig:conv_H3S}
		The convergence test of EPC parameters $\lambda(m)$ with respect to the number of beads in H3S.}
\end{figure}

For solid deuterium, we perform PIMD simulations in a $3\times3\times3$ supercell at $250 K$, and calculate EPC parameters on a $18\times18\times18$ $k$-grid of the supercell. 
For solid H$_3$S, we perform PIMD simulations in a $3\times3\times3$ supercell at $190 K$, and calculate EPC parameters on a $8\times8\times8$ $k$-grid of the supercell. 
The predicted $T_c$ with respect to the number of beads are shown in Fig.~\ref{fig:conv}. In Figs.\ref{fig:conv_D} and \ref{fig:conv_H3S}, we also show convergence tests of EPC parameters $\lambda(m)$. We find that $N_b=16$ yields relevant convergence in both cases.  In the main text, we use $N_b=24$ and $N_b=16$ in Deuterium and H$_3$S, respectively, which yield convergence within $\Delta T_c\approx1K$.

\bibliographystyle{apsrev4-2}
\bibliography{Reference}

\end{document}